\documentclass[10pt,a4paper]{article}
\usepackage{graphicx}

\newcommand{\be}{\begin{equation}}
\newcommand{\ee}{\end{equation}}
\newcommand{\bea}{\begin{eqnarray}}
\newcommand{\eea}{\end{eqnarray}}

\begin{document}

\title{The role of squeezing in quantum key distribution based on
homodyne detection and post-selection}

\author{Peter Horak\\
Optoelectronics Research Centre, University of Southampton,\\
Southampton SO17 1BJ, UK}

\date{\today}

\maketitle

\begin{abstract}
The role of squeezing in quantum key distribution with continuous
variables based on homodyne detection and post-selection is
investigated for several specific eavesdropping attacks. It is
shown that amplitude squeezing creates strong correlations between
the signals of the legitimate receiver and a potential
eavesdropper. Post-selection of the received pulses can therefore
be used to reduce the eavesdropper's knowledge of the raw key,
which increases the secret key rate by orders of magnitude over
large distances even for modest amounts of squeezing.
\end{abstract}


\section{Introduction}

Quantum cryptography or quantum key distribution was first
proposed nearly two decades ago \cite{bb84} as a method to exploit
properties of small quantum systems to establish a secret key
between a sender (Alice) and a receiver (Bob) which is provably
secure against attacks by an eavesdropper (Eve). In recent years
this field has attracted a lot of research and rapid progress has
been achieved in the theoretical description and in experimental
demonstrations. In particular, the range over which a secure key
can be established has now been extended up to 101 km in optical
fibres \cite{toshiba,mitsubishi,gisin} and 23 km in air
\cite{weinfurter,hughes}. However, most of these systems are based
on the transmission of single photons and suffer from inefficient
and slow single-photon sources and detectors. These technical
restrictions lead to very low secret key bit rates, e.g.\ down to
15 bit/s in the case of \cite{toshiba}, and to a maximum distance
beyond which no secure key can be transmitted at all
\cite{norbert2}.

In order to overcome these limitations, a new class of quantum
crypto-systems has been proposed, where information is encoded in
continuous variables such as field quadratures
\cite{ralph,hillery,reid,gottesmann}, photon numbers \cite{funk},
or polarisation \cite{leuchspol}. These schemes are based on the
transmission of non-classical states of light, i.e.\ squeezed or
entangled states, and are restricted to transmission lines with
less than 50 \% loss.

Very recently, however, it has been shown that quantum key
distribution can operate over arbitrarily lossy channels using
only classical, i.e.\ coherent, quantum states of light. This is
achieved by reverse reconciliation techniques \cite{revrecon} or
by post-selection \cite{leuchs,namiki}. The first quantum
crypto-system of this kind has already been demonstrated by
Grangier and co-workers \cite{grangier}.

In this paper, the effect of squeezing on the performance of
quantum key distribution with post-selection is investigated. For
the schemes operating with transmission lines with less than 50\%
loss \cite{ralph,hillery,reid,gottesmann,funk,leuchspol},
squeezing guarantees that the smaller part of the pulse lost in
the transmission or intercepted by Eve has much larger quantum
uncertainties and thus less information than the larger part
received by Bob. One therefore might think that for losses larger
than 50\%, where Eve potentially has access to a larger fraction
of the pulse than Bob, squeezing might act in favour of the
eavesdropper. The main purpose of this work is to show that
squeezing still is beneficial beyond the 50\% loss limit in
post-selection based schemes.

This work is organised as follows. Section \ref{sec:scheme}
briefly reviews the quantum key distribution scheme \cite{namiki}.
Then two classes of eavesdropping attacks are investigated which
clearly demonstrate the merits of using squeezed states. First,
intercept-resend strategies are analysed in section
\ref{sec:resend}, where Eve performs a measurement on the whole
pulse sent by Alice and resends a pulse to Bob according to her
measurement results. Two specific measurements based on quadrature
detection and phase detection, respectively, are discussed as well
as the optimal orthogonal state projection measurement. It is
shown that squeezing reduces Eve's chances of success with this
attack. The second attack, discussed in section \ref{sec:sup}, is
a superior channel attack, where Eve keeps that part of the pulse
which would normally be lost in the transmission channel from
Alice to Bob, and transmits the rest of the pulse through a
lossless channel to Bob. This kind of attack cannot be detected by
Alice and Bob. The maximum amount of secret information which can
be obtained despite Eve's presence is calculated for an Eve
restricted to quadrature measurements and a lower bound is given
for the case where Eve is allowed to perform arbitrary
measurements. Also for this kind of attack squeezing can
significantly enhance the secret key rate. Finally, the results
are summarised in section \ref{sec:conclusions}.


\section{Quantum key distribution with homodyne detection and post-selection}
\label{sec:scheme}

The scheme presented here is a generalisation of the recent
proposal by Namiki and Hirano \cite{namiki}, where Alice sends
squeezed states instead of coherent (classical) pulses. Without
the post-selection process the scheme is also equivalent to the
scheme by Funk and Raymer \cite{funk} which is based on the
simultaneous transmission of two pulses in orthogonal polarisation
states obtained from a parametric amplification process. It can be
shown by a unitary transformation that the latter entangled state
is equivalent to a product state of a weak amplitude-squeezed
pulse (corresponding to the pulse sent by Alice here) and a strong
(phase-squeezed) pulse (the local oscillator used for homodyne
detection).

The scheme works as follows. First, Alice prepares a
minimum-uncertainty squeezed pulse with field amplitude $\alpha_0$
and squeezing parameter $r$. Both $\alpha_0$ and $r$ are assumed
to be real numbers, $r>0$ corresponds to an amplitude-squeezed
state, $r<0$ to a phase-squeezed state, and $r=0$ to a coherent
state. Alice then applies a phase shift $\theta\in
\{0,\pi/2,\pi,3\pi/2\}$ to the pulse, where $\theta=0$
($\theta=\pi/2$) is interpreted as a bit 1 in basis 1 (basis 2)
and $\theta=\pi$ ($\theta=3\pi/2$) as a bit 0 in basis 1 (2). The
resulting state reads
 \be
 |\psi_\theta\rangle =
 \exp\left(\alpha_0 e^{i\theta}a^\dagger-\alpha_0 e^{-i\theta}a\right)
   \exp\left((r e^{-2i\theta}a^2-r e^{2i\theta}a^{\dagger2})/2\right) |0\rangle
 \ee
where $a$ ($a^\dagger$) is Alice's photon annihilation (creation)
operator and $|0\rangle$ is the vacuum state.

Next, Alice sends this state to Bob. During this transmission the
pulse experiences attenuation, that is, a fraction of the pulse is
either absorbed by the environment or intercepted by Eve. In both
cases the transmission is modeled by a beam-splitter of amplitude
transmission $T$ and reflection $R$, where $T^2+R^2=1$. The output
mode operators $(b,b^\dagger)$ and $(e,e^\dagger)$ correspond to
the modes in Bob's and Eve's detectors, respectively, and are
related to the input mode operators $(a,a^\dagger)$ and the
additional (vacuum) input mode $(v,v^\dagger)$ of the
beam-splitter by
 \be
 \left(\begin{array}{c} b \\ e \end{array}\right) =
  \left(\begin{array}{cc} T & R \\ -R & T \end{array}\right)
  \left(\begin{array}{c} a \\ v \end{array}\right).
 \label{eq:split}
 \ee
For the following it is convenient to express the quantum state of
the transmitted pulse $|\psi_\theta\rangle$ in terms of its Wigner
quasi-probability function for Bob's (complex) measurement
variable $\beta$ and Eve's variable $\epsilon$. One finds
 \bea
 W_\theta(\beta,\epsilon) & = & \frac{4}{\pi^2}
   \exp\left\{-2\left[
          e^{2r} {\cal R}\left\{\alpha \right\}^2+
          e^{-2r}{\cal I}\left\{\alpha \right\}^2
       \right] \right\} \nonumber\\
   & &\times \exp\left\{-2\left[
          \left|R\beta+T\epsilon\right|^2
       \right] \right\}
 \eea
where ${\cal R}$ and ${\cal I}$ denote real and imaginary parts,
respectively, and
 \be
 \alpha = (T\beta-R\epsilon)e^{-i\theta}-\alpha_0.
 \ee

In the next step of the quantum key distribution protocol, Bob
randomly decides which of the two basis sets he uses for his
detection. If he chooses basis 1 he measures the real part
$\beta_r$ of his variable $\beta$ by homodyne detection, otherwise
he measures the imaginary part $\beta_i$. His measurement results
follow the probability distributions
 \bea
 P_\theta(\beta_r) & = & \int W_\theta(\beta,\epsilon)
    d\beta_i d\epsilon_r d\epsilon_i, \\
 P_\theta(\beta_i) & = & \int W_\theta(\beta,\epsilon)
    d\beta_r d\epsilon_r d\epsilon_i.
 \eea
If his measurement result fulfills $|\beta_{r,i}|<\beta_c$, where
$\beta_c$ is a fixed threshold value, Bob tells Alice to disregard
this bit. Otherwise he assigns value 1 to his bit if
$\beta_{r,i}>\beta_c$, or value 0 if $\beta_{r,i}<-\beta_c$. This
post-selection is the key to allow for quantum key distribution
over, in principle, arbitrarily large distances
\cite{leuchs,namiki}. Alice and Bob additionally announce their
used basis sets and disregard all bits where they used different
sets.

After this step, Alice and Bob share a string of bits which in
general contains some errors and about which Eve has a certain
amount of information. Alice and Bob then use (classical) error
correction and privacy amplification protocols which leaves them
with an identical secret key about which Eve has only negligible
information.

As an example let us assume that Alice sends the state
$|\psi_0\rangle$, i.e.\ $\theta=0$. In this case, Bob's
probabilities read
 \bea
 P_0(\beta_r) & = & \sqrt{\frac{2}{\pi}}
  \frac{\exp\left\{-2\frac{(\beta_r-T\alpha_0)^2}{T^2 e^{-2r}+R^2}\right\}}
  {\sqrt{T^2 e^{-2r}+R^2}}, \label{eq:wbr}\\
 P_0(\beta_i) & = & \sqrt{\frac{2}{\pi}}
  \frac{\exp\left\{-2\frac{\beta_i^2}{T^2 e^{2r}+R^2}\right\}}
  {\sqrt{T^2 e^{2r}+R^2}}. \label{eq:wbi}
 \eea
For an amplitude-squeezed state ($r>0$), Bob thus finds a Gaussian
distribution with a narrow width if he measures the quadrature
$\beta_r$, and a broad width for measurements of $\beta_i$. The
probability of finding a bit 1 or 0 in the correct basis (basis 1
in the example here) is thus
 \bea
 P(1) & = & \int^{\infty}_{\beta_c} P_0(\beta_r) d\beta_r
  = \frac{1}{2}+\frac{1}{2}
  \Phi\left(\frac{\sqrt{2}(T\alpha_0-\beta_c)}{\sqrt{T^2
  e^{-2r}+R^2}}\right), \label{eq:p1}\\
 P(0) & = & \int_{-\infty}^{-\beta_c} P_0(\beta_r) d\beta_r
  = \frac{1}{2}-\frac{1}{2}
  \Phi\left(\frac{\sqrt{2}(T\alpha_0+\beta_c)}{\sqrt{T^2
  e^{-2r}+R^2}}\right), \label{eq:p0}
 \eea
where $\Phi$ denotes the usual error function. For $r=0$ equations
(\ref{eq:wbr})-(\ref{eq:p0}) reduce to the results found in
\cite{namiki}. The fraction of accepted bits is
 \be
 r_{acc} = \frac{P(1)+P(0)}{2},
 \label{eq:racc}
 \ee
where the factor 1/2 comes from disregarding all bits where Alice
and Bob use different basis sets, and the bit error rate of
accepted bits is
 \be
 \delta = \frac{P(0)}{P(0)+P(1)}.
 \ee
The expected amount of Shannon information shared by Alice and Bob
per accepted bit is obtained by averaging over Bob's measurement
outcomes,
 \bea
 I_{AB} & = & \int_{\beta_c}^\infty d\beta_r
   \frac{P_0(\beta_r)+P_0(-\beta_r)}{P(0)+P(1)} \nonumber\\
   & & \times \left\{1+\delta(\beta_r)\log_2\delta(\beta_r)
       +(1-\delta(\beta_r))\log_2(1-\delta(\beta_r))\right\}
 \label{eq:iab}
 \eea
where
 \be
 \delta(\beta_r) = \frac{P_0(-\beta_r)}{P_0(\beta_r)+P_0(-\beta_r)}
 \label{eq:dbr}
 \ee
is the bit error rate conditioned on Bob's measured value of
$\beta_r$. Thus the average information gain per transmitted bit
is
 \be
 G_{AB} = r_{acc}I_{AB}.
 \label{eq:gab}
 \ee

\begin{figure}[tb]
 \centerline{
 \includegraphics[height=4.5cm]{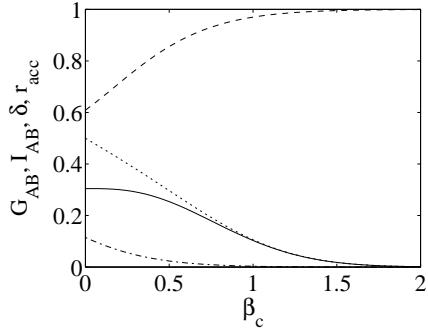}
 }
 \caption{Quantum key distribution over a lossless channel:
 acceptance rate $r_{acc}$ (dotted line), bit error rate $\delta$ per accepted bit
 (dash-dotted), information per accepted bit $I_{AB}$ (dashed) and
 average information gain per transmitted bit $G_{AB}$ (solid
 line). The parameters are $\alpha_0=0.6$, $r=0$, $T=1$.
 }
 \label{fig:lossless}
\end{figure}

An example for the quantities (\ref{eq:racc})-(\ref{eq:gab}) as a
function of the threshold value $\beta_c$ is shown in figure
\ref{fig:lossless}. For the chosen parameters and for $\beta_c=0$,
half of all transmitted bits are accepted, but the bit error rate
is relatively large and thus the information $I_{AB}$ per accepted
bit is low. Increasing $\beta_c$ reduces $r_{acc}$ but also
decreases the bit error rate. The figure shows that the average
information gain $G_{AB}$ per bit decreases with increasing
$\beta_c$. It therefore seems advantageous for Bob to set
$\beta_c=0$. However, in a realistic setup Alice and Bob will try
to optimise the \textit{secret} key bit rate, i.e.\ the difference
between their shared information and the information leaked to
Eve. This subject will be discussed later in section
\ref{sec:sup}. Note that due to the form of the expressions
(\ref{eq:p1}), (\ref{eq:p0}) the qualitative behaviour of the
quantities plotted in figure \ref{fig:lossless} is the same for
all squeezing parameters $r$ and transmission coefficients $T$.


\section{Intercept-resend attacks}
\label{sec:resend}

In this section I will investigate the security of the protocol
against eavesdropping attacks where Eve intercepts each pulse,
performs a measurement on it, and then sends a pulse to Bob in the
state which she thinks was initially prepared by Alice. First, two
specific eavesdropping strategies based on simultaneous
measurements of both quadrature components and on phase
measurement, respectively, will be analysed. In the third
subsection the ideal intercept-resend attack will be constructed
based on general quantum measurement theory.

In the following I will concentrate on Eve's probability
$p_{corr}$ to guess correctly the bit and the basis of Alice's
pulse. With probability $1-p_{corr}$, Eve will thus transmit a
pulse to Bob that differs from the original one, which Alice and
Bob can detect by an increase of the bit error rate $\delta$ or,
for example, by a change of the distribution of Bob's measured
quadrature components or the measured value of $r_{acc}$.

\subsection{Simultaneous quadrature measurement}

For the first attack investigated here let us assume that Eve
splits each pulse into two parts on a 50-50 beam-splitter,
measures one quadrature component in one output arm, and
simultaneously the orthogonal quadrature component in the other
arm.

Eve's measurement results are described by the expressions derived
in section \ref{sec:scheme} if $T=R=1/\sqrt{2}$ and assuming that
Eve performs measurements on both output arms of the beam-splitter
(\ref{eq:split}). Specifically, let us assume that Eve measures
$\beta_r$ and $\epsilon_i$. The joint probability distribution is
given by
 \be
 P_\theta(\beta_r,\epsilon_i) = \int W_\theta(\beta,\epsilon)
    d\beta_i d\epsilon_r
 \ee
which for all parameters factorises into two independent
probability distributions
 \be
 P_\theta(\beta_r,\epsilon_i) =
 P_\theta(\beta_r)P_\theta(\epsilon_i).
 \ee
For $\theta=0$, $P_\theta(\beta_r)$ is given by equation
(\ref{eq:wbr}) and
 \be
 P_0(\epsilon_i) = \sqrt{\frac{2}{\pi}}
  \frac{\exp\left\{-2\frac{\epsilon_i^2}{T^2+R^2 e^{2r}}\right\}}
  {\sqrt{T^2+R^2 e^{2r}}} \label{eq:wei}
 \ee
with $T^2=R^2=1/2$.

If Eve's measurement outcome is $(\beta_r,\epsilon_i)$ she will
attribute the pulse to the value $\theta\in
\{0,\pi/2,\pi,3\pi/2\}$ for which $P_\theta(\beta_r,\epsilon_i)$
is maximum. For $r\leq 0$ this yields
 $\theta=0$ for $\beta_r\geq|\epsilon_i|$,
 $\theta=\pi/2$ for $\epsilon_i>|\beta_r|$,
 $\theta=\pi$ for $-\beta_r\geq|\epsilon_i|$,
 $\theta=3\pi/2$ for $-\epsilon_i>|\beta_r|$.
For $r>0$, which is the interesting case for the purpose of this
work, finding the most probable value of $\theta$ is more
complicated. Let us assume that Eve finds $\beta_r,\epsilon_i>0$.
She will then attribute this result to $\theta=0$ if
 \be
 \epsilon_i < \beta_r<\frac{\sqrt{2}\alpha_0}{1-e^{-2r}}-\epsilon_i
 \mbox{\quad or \quad}
 \frac{\sqrt{2}\alpha_0}{1-e^{-2r}}-\epsilon_i<\beta_r<\epsilon_i.
 \label{eq:area}
 \ee
The remaining areas in the quadrant $\beta_r,\epsilon_i>0$ are
attributed to $\theta=\pi/2$, and the results for the other three
quadrants of $(\beta_r,\epsilon_i)$ follow by symmetry. Eve's rate
of success for this attack is thus given by
 \be
 p_{corr} = 2\int_{A_0} P_0(\beta_r,\epsilon_i) d\beta_r
 d\epsilon_i
 \ee
where $A_0$ is the area defined by (\ref{eq:area}).

\begin{figure}[tb]
 \centerline{
 \includegraphics[width=5.5cm]{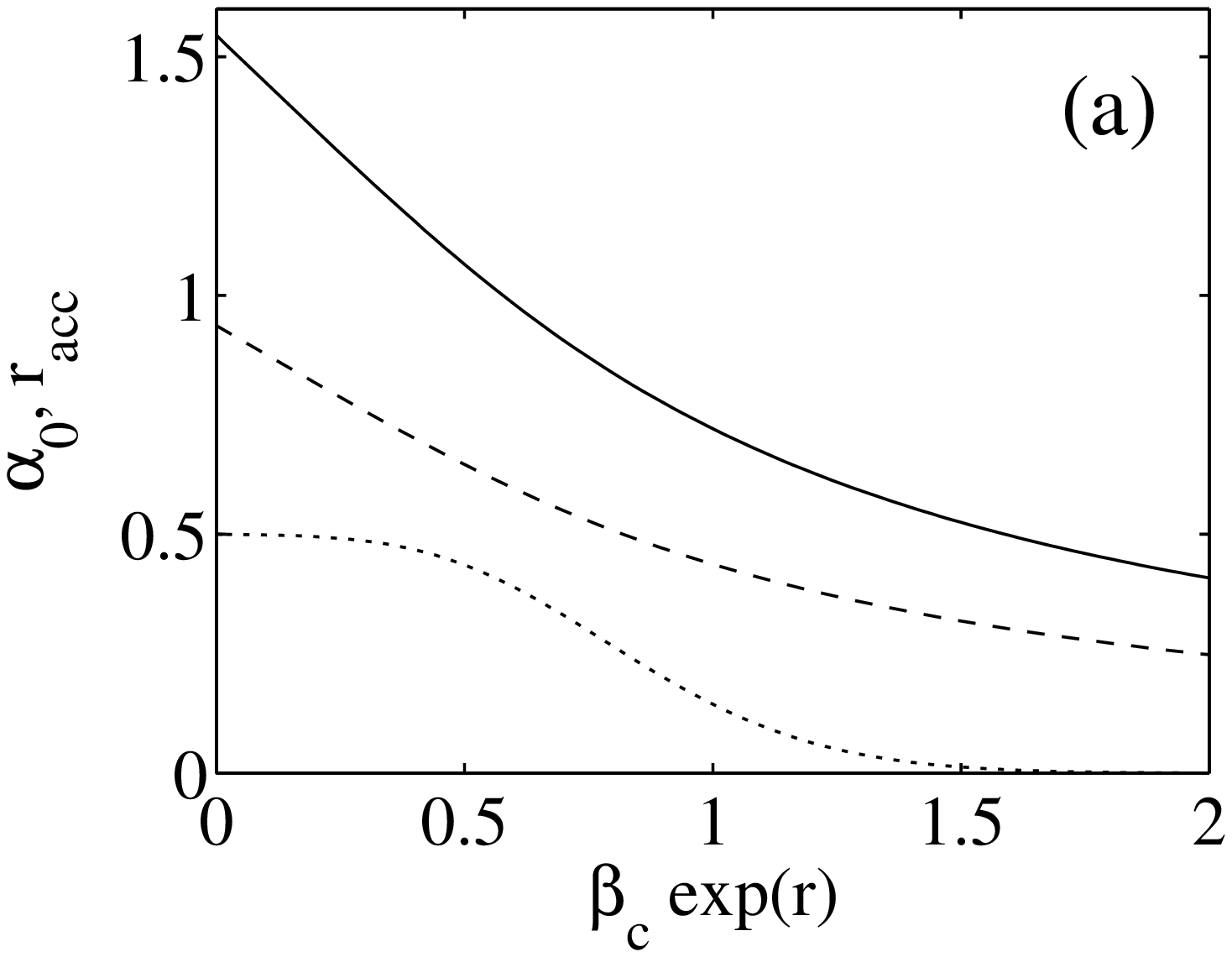}
 \includegraphics[width=5.5cm]{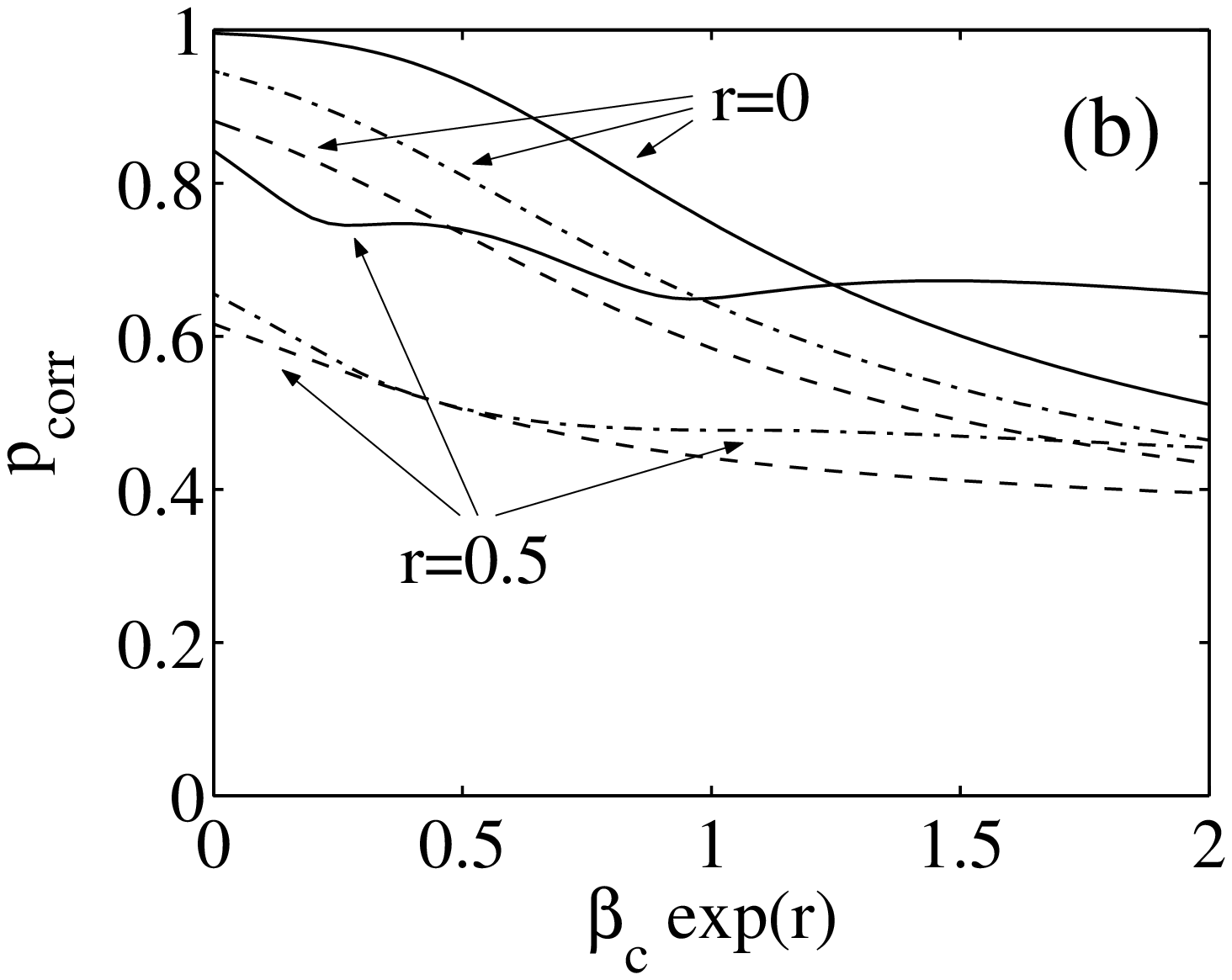}
 }
 \caption{Intercept-resend attack for different
 post-selection thresholds $\beta_c$ and varying squeezing. (a)
 Choice of $\alpha_0$ for $r=0$ (solid line) and $r=0.5$ (dashed)
 such that $\delta=10^{-3}$. Dotted line is the corresponding
 fraction $r_{acc}$ of accepted bits. (b) Eve's success rate
 $p_{corr}$ for simultaneous quadrature measurements (dashed
 lines), phase measurements (dash-dotted), and optimised state
 projection measurements (solid).
 }
 \label{fig:spliterr}
\end{figure}

Figure \ref{fig:spliterr} shows the efficiency of the simultaneous
measurement attack for various parameters. For any given
post-selection threshold $\beta_c$ and squeezing parameter $r$,
Alice adjusts the mean amplitude $\alpha_0$ such that
$\delta=10^{-3}$ for lossless transmission and without Eve's
interference. The values for $\alpha_0$ for $r=0$ and $r=0.5$ are
shown in figure \ref{fig:spliterr}(a) together with the resulting
fraction of accepted bits $r_{acc}$ (the abscissa of the plot is
scaled such that this curve is identical for all values of $r$).
The dashed lines in figure \ref{fig:spliterr}(b) show Eve's
probability $p_{corr}$ of success for $r=0$ and $r=0.5$. For any
amount of squeezing, Bob can decrease Eve's success rate by
increasing $\beta_c$. The reason for this is that larger
post-selection thresholds allow for the use of weaker pulses, see
figure (a), which in turn gives larger overlaps of the pulses
corresponding to different values of $\theta$ in Eve's
measurement. However, larger thresholds also lead to lower bit
acceptance rates and thus slower secret key generation. The use of
squeezed pulses helps to reduce this problem, since in this case
already low values of $\beta_c$ lead to a low probability of
success for Eve's attack. This is due to the lower intensity of
the pulses sent by Alice and the increased uncertainty in the
quadrature components measured by Eve because of the vacuum noise
on the second input port of Eve's beam-splitter.

\subsection{Phase measurement}

Using simultaneous quadrature measurement, Eve obtains estimates
for the amplitude and phase of each pulse. However, in the scheme
discussed here the amplitude of all states $|\psi_{\theta}\rangle$
is the same and Eve is in fact only interested in measuring the
phase. A better eavesdropping attack might therefore be based on
an adaptive phase measurement \cite{wiseman} which uses heterodyne
detection with nonlinear feedback and which in certain limits can
approach the optimal measurement of the phase of a single pulse.

The distribution of the phase $\phi\in [0,2\pi)$ detected by an
optimal phase measurement on a state $|\psi_{\theta}\rangle$ is
given by
 \be
 P_\theta(\phi) = \frac{1}{2\pi}
    \left|\langle\phi|\psi_\theta\rangle\right|^2
 \label{eq:pphi}
 \ee
where
 \be
 |\phi\rangle = \sum_{n=0}^{\infty} e^{i n\phi}
    \frac{a^{\dagger n}}{\sqrt{n!}} |0\rangle.
 \ee
If Eve measures a certain phase $\phi$, she will attribute this
result to that state $|\psi_\theta\rangle$ for which
$P_\theta(\phi)$ is maximum. The success rate of this
eavesdropping strategy is thus
 \be
 p_{corr} = \int_{F_0} P_0(\phi)\, d\phi
 \label{eq:pcphi}
 \ee
where
 \be
 F_0 = \left\{\phi | P_0(\phi)>
   \max(P_{\pi/2}(\phi),P_{\pi}(\phi),P_{3\pi/2}(\phi)) \right\}.
 \ee

Evaluation of equation (\ref{eq:pphi}) in the Fock basis and
subsequent numerical integration of (\ref{eq:pcphi}) yield the
results shown by the dash-dotted curves in figure
\ref{fig:spliterr}. Compared to the simultaneous quadrature
measurement discussed above the phase measurement leads to a
larger success rate $p_{corr}$ for Eve. However, the qualitative
behaviour is similar. In particular, squeezing is found to reduce
the eavesdropping efficiency significantly also for phase
measurements.

\subsection{Orthogonal state projection}
\label{sec:interceptideal}

In the scheme presented here, each pulse sent by Alice is one of
the four linearly independent and non-orthogonal states
$|\psi_\theta\rangle$. For this particular case, Eve's ideal
measurement, i.e.\ the one which gives the largest value of
$p_{corr}$, is known to be a projection measurement on four
orthogonal states in the Hilbert space ${\cal H}_s$ spanned by the
four states $|\psi_\theta\rangle$ \cite{ban}. Such a projection
measurement is extremely difficult to realise experimentally, but
theoretical proposals for situations involving only two coherent
states exist, see e.g.\ \cite{sasaki}.

In the following I will construct the optimal projection
measurement. For this it is necessary to find four orthogonal
states $|\varphi_\theta\rangle$, $\theta\in
\{0,\pi/2,\pi,3\pi/2\}$, which maximise
 \be
 p_{corr}=\frac{1}{4}\sum_{\theta}
    |\langle\varphi_\theta|\psi_\theta\rangle|^2.
 \ee
Since the states $|\psi_\theta\rangle$ are obtained from a single
state by the phase shift operator $U=\exp(i\pi a^\dagger a/2)$, it
can be assumed that the states $|\varphi_\theta\rangle$ obey the
same symmetry, that is,
 \be
 |\varphi_{\theta+\pi/2}\rangle = U |\varphi_\theta\rangle.
 \label{eq:phisym}
 \ee
The operator $U$ has the eigenvalues $\pm 1$ and $\pm i$. Let
$|u_\nu\rangle$, $\nu\in\{\pm 1,\pm i\}$, denote the corresponding
ortho-normal eigenvectors in ${\cal H}_s$. Then, the states
$|\varphi_\theta\rangle$ can be written as
 \be
 |\varphi_\theta\rangle=\sum_\nu q_{\theta,\nu} |u_\nu\rangle.
 \ee
The orthogonality and normalisation of the states
$|\varphi_\theta\rangle$ together with equation (\ref{eq:phisym})
implies $|q_{\theta,\nu}|=1/2$. In order to maximise
$|\langle\varphi_\theta|\psi_\theta\rangle|$ and therefore
$p_{corr}$ the complex phases of $q_{\theta,\nu}$ are obtained (up
to a global phase) as
 \be
 q_{\theta,\nu} = \frac{1}{2}
    \frac{\langle u_\nu|\psi_\theta\rangle}
    {|\langle u_\nu|\psi_\theta\rangle|}.
 \ee
The states $|u_\nu\rangle$, $|\varphi_\theta\rangle$ and therefore
the optimal value of $p_{corr}$ can be obtained numerically for
given parameters $\alpha_0$ and $r$. It has also been checked that
the set of states $|\varphi_\theta\rangle$ constructed in this way
yields the optimal orthogonal projection measurement even without
the a priori requirement of the symmetry (\ref{eq:phisym}).

Results of this optimal intercept-resend attack for two values of
$r$ are shown by the solid curves in figure \ref{fig:spliterr}(b).
As expected, Eve's rate of success $p_{corr}$ is larger for this
attack than for either of the two attacks analysed above for all
parameters. Increasing the post-selection threshold $\beta_c$
reduces $p_{corr}$ to below 70\% for coherent as well as squeezed
states for the parameters shown in the figure. Using squeezed
states allows to use lower thresholds and therefore to obtain
larger acceptance rates $r_{acc}$.

However, it is important to note that for too small values of
$\alpha_0$, corresponding to larger $\beta_c$ in figure
\ref{fig:spliterr}, squeezing \textit{increases} $p_{corr}$ and
thus helps the eavesdropper. This can be understood in the
following way. Without squeezing ($r=0$) and in the limit of
$\alpha_0\rightarrow 0$, all four states $|\psi_\theta\rangle$
become identical to the vacuum state. Eve's probability of
guessing bit and basis correctly thus approaches 1/4. For squeezed
states, on the other hand, $|\psi_0\rangle$ and $|\psi_\pi\rangle$
approach the vacuum state squeezed by $r$ in the $a+a^\dagger$
quadrature, but $|\psi_{\pi/2}\rangle$ and $|\psi_{3\pi/2}\rangle$
approach the vacuum state squeezed by $r$ in the $a-a^\dagger$
quadrature. Thus, Eve can still detect the basis (but not the bit)
in which a pulse was sent with large probability. For example for
$r=0.5$ this gives $p_{corr}\approx 0.4$.


\section{Superior channel attack}
\label{sec:sup}

In the second class of eavesdropping attacks discussed in this
paper, Eve is assumed to possess an ideal quantum memory and a
lossless channel. She uses a beam-splitter to extract that part of
the pulse which would normally be lost in the transmission channel
from Alice to Bob, stores it in her quantum memory and sends the
rest of the pulse through the lossless channel to Bob. She then
waits for Bob's public announcement which basis he used and
afterwards she performs a measurement on her part of the pulse.
This kind of attack cannot be detected by Alice and Bob since
Eve's interference is indistinguishable from the standard
transmission loss.

Since Alice and Bob cannot infer Eve's presence from their data,
they have to assume the worst case and therefore attribute all of
their transmission losses to Eve. In order to establish a secure
key it is necessary to derive an upper bound of the information
available to Eve, such that in the final stage of the protocol
Eve's knowledge about the key can be reduced to arbitrarily low
levels using privacy amplification. Here I will derive such an
upper bound following the lines of L\"utkenhaus
\cite{norbert1,norbert2}.


\subsection{Quadrature measurement}

In a first step I will assume that Eve uses only linear optics for
her measurement, that is, after Bob's public announcement of his
basis Eve uses a balanced homodyne detector to measure the correct
quadrature component of her part of the pulse.

\begin{figure}[tb]
 \centerline{
 \includegraphics[height=4.5cm]{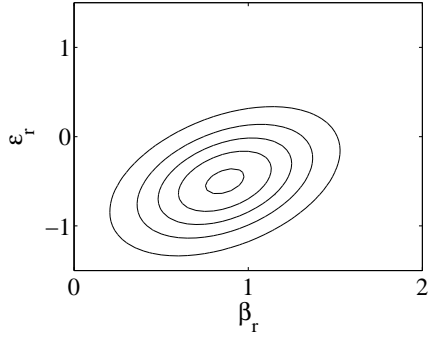}
 }
 \caption{Contour plot of the joint probability distribution
 $P_0(\beta_r,\epsilon_r)$ with $\alpha_0=1$,
 $r=0.5$, $T^2=0.75$. The maximum of $P_0(\beta_r,\epsilon_r)$
 occurs at $\beta_r=T\alpha_0=\sqrt{3}/2$,
 $\epsilon_r=-R\alpha_0=-1/2$.
 }
 \label{fig:correlation}
\end{figure}

The fundamental quantity of the following derivation is the joint
measurement probability $P_\theta(\beta_r,\epsilon_r)$
 \be
 P_\theta(\beta_r,\epsilon_r) = \int W_\theta(\beta,\epsilon)
    d\beta_i d\epsilon_i
 \ee
that Bob obtains the value $\beta_r$ and simultaneously Eve
obtains $\epsilon_r$ when Alice sends a pulse
$|\psi_\theta\rangle$. For $\theta=0$ this can be evaluated as
 \be
 P_0(\beta_r,\epsilon_r)=\frac{2}{\pi}e^r\exp\left\{
    -2\left[e^{2r}(T\beta_r-R\epsilon_r-\alpha_0)^2
            +(R\beta_r+T\epsilon_r)^2\right]
   \right\}.
 \label{eq:entangle}
 \ee
For coherent states, $r=0$, this can be factorised and hence Bob's
and Eve's measurements are independent. For squeezed states, on
the other hand, the joint probability distribution does not
factorise and the two measurements are correlated. In particular,
for $r>0$ as depicted in figure \ref{fig:correlation}, the
measurements are anti-correlated, that is, if Bob measures a
relatively large value of $\beta_r$, Eve is likely to measure a
small value of $\epsilon_r$. Hence, by choosing a larger
post-selection threshold Bob can increase Eve's bit error rate and
thus reduce Eve's information.

Assuming ideal error correction and privacy amplification, a lower
bound for the gain of secret information per transmitted pulse is
given by \cite{norbert1,norbert2}
 \be
 S_{AB} = r_{acc}(I_{AB}-\tau)
 \label{eq:sab}
 \ee
where $r_{acc}$ is given by (\ref{eq:racc}), $I_{AB}$ by
(\ref{eq:iab}), and the fraction $\tau$ by which the raw key is
shortened during privacy amplification is
 \be
 \tau = 1+\log_2 P_c.
 \label{eq:tau}
 \ee
Here, $P_c$ is the collision probability given by \cite{norbert1}
 \be
 P_c = \frac{1}{2}\int d\epsilon_r
   \frac{P_0(\epsilon_r|\beta_c<|\beta_r|)^2+P_\pi(\epsilon_r|\beta_c<|\beta_r|)^2}
        {P_0(\epsilon_r|\beta_c<|\beta_r|)+P_\pi(\epsilon_r|\beta_c<|\beta_r|)}
 \label{eq:pcoll}
 \ee
and
 \be
 P_\theta(\epsilon_r|\beta_c<|\beta_r|) =
   \int_{\beta_c<|\beta_r|}d\beta_r
   \frac{P_\theta(\beta_r,\epsilon_r)}{P(0)+P(1)}
 \label{eq:pcond}
 \ee
is Eve's probability distribution under the condition that a pulse
$\theta$ was sent and that Bob accepted the bit in his
post-selection. Equation (\ref{eq:pcond}) can be evaluated
analytically, yielding
 \bea
 & &P_0(\epsilon_r|\beta_c<|\beta_r|) =
 \frac{P_0(\epsilon_r)}{P(0)+P(1)}
 \bigg\{1 \nonumber\\
 & &\quad\quad
  -\frac{1}{2}\Phi\left(\sqrt{2}
     \frac{\beta_c(R^2+T^2e^{2r})+\epsilon_r TR(1-e^{2r})-\alpha_0Te^{2r}}
     {\sqrt{R^2+T^2e^{2r}}}\right) \nonumber\\
  & &\quad\quad
  -\frac{1}{2}\Phi\left(\sqrt{2}
     \frac{\beta_c(R^2+T^2e^{2r})-\epsilon_r TR(1-e^{2r})+\alpha_0Te^{2r}}
     {\sqrt{R^2+T^2e^{2r}}}\right)
 \bigg\}
 \eea
where
 \be
 P_0(\epsilon_r) = \sqrt{\frac{2}{\pi}}
  \frac{\exp\left\{-2\frac{(\epsilon_r+R\alpha_0)^2}{T^2+R^2 e^{-2r}}\right\}}
  {\sqrt{T^2+R^2 e^{-2r}}}.
 \ee
For $r=0$, this simplifies to
$P_0(\epsilon_r|\beta_c<|\beta_r|)=P_0(\epsilon_r)$ due to the
independence of Bob's and Eve's measurements. Equations
(\ref{eq:iab}) and (\ref{eq:pcoll}) must be calculated numerically
in order to get the secret key bit rate $S_{AB}$, equation
(\ref{eq:sab}).

\begin{figure}[tb]
 \centerline{
 \includegraphics[width=6cm]{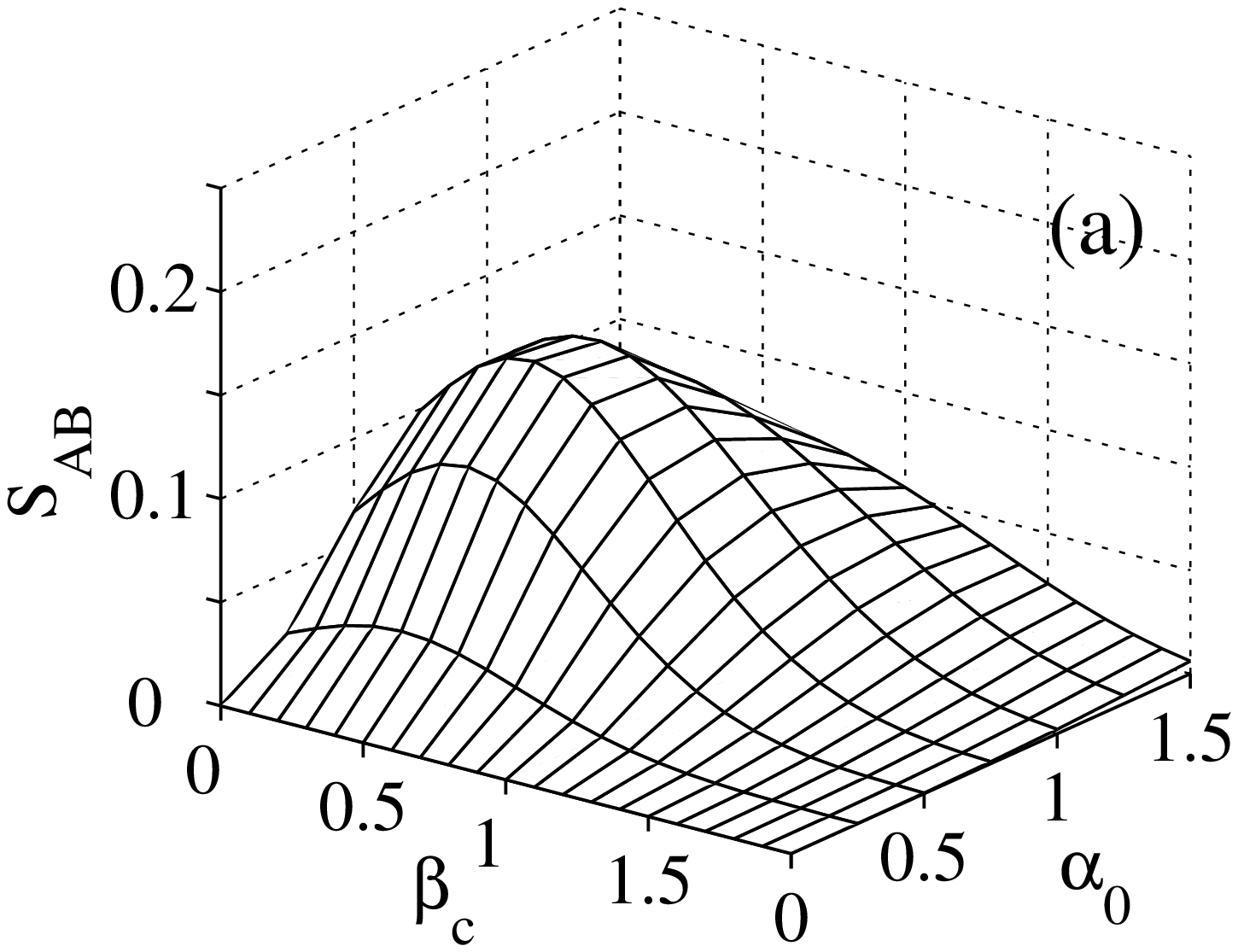}
 \includegraphics[width=6cm]{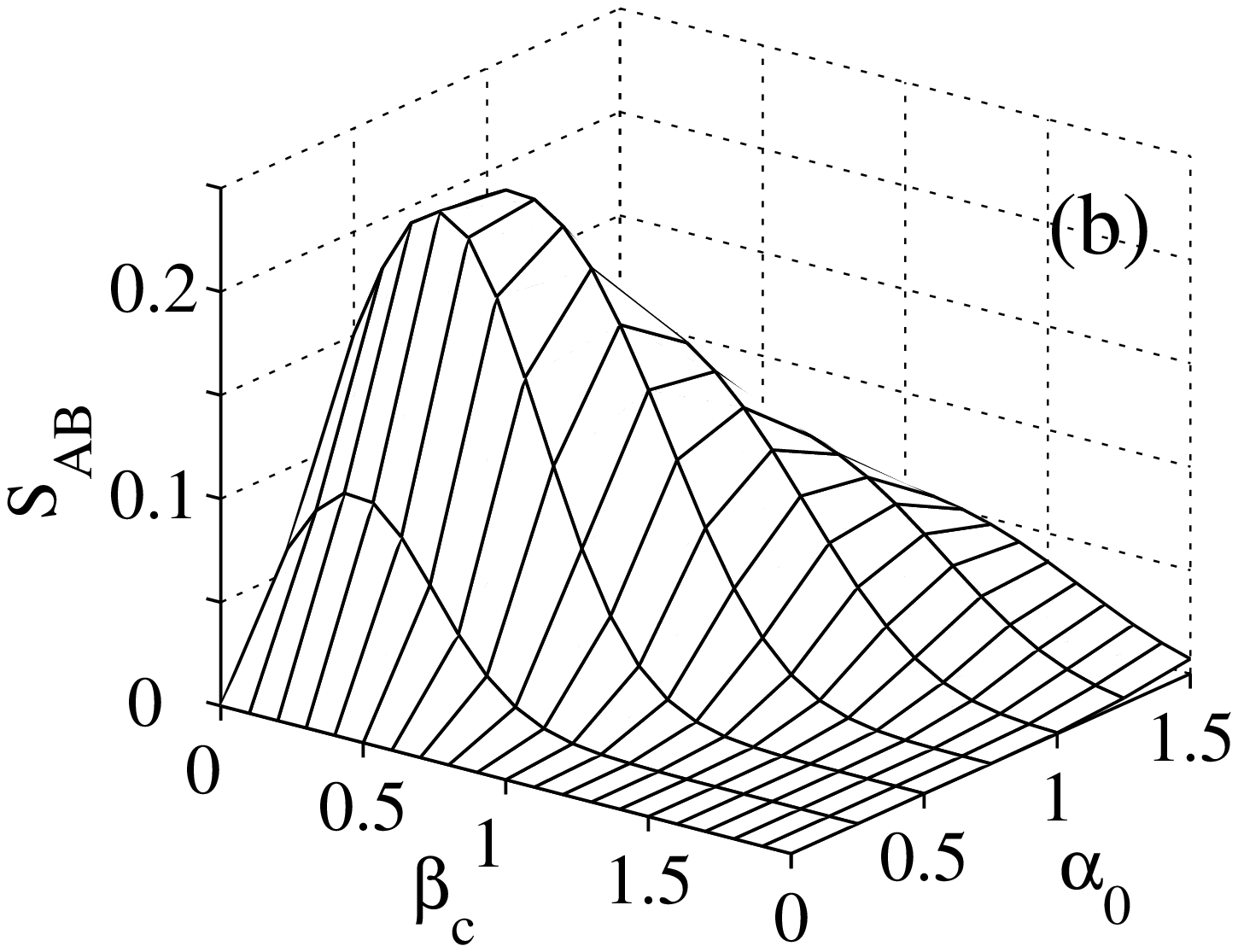}
 }
 \caption{Secret key bit rate $S_{AB}$ versus mean pulse amplitude
 $\alpha_0$ and post-selection threshold $\beta_c$ for (a)
 coherent pulses and (b) squeezed pulses with $r=0.5$.
 Channel transmission is $T^2=0.75$.
 }
 \label{fig:secure}
\end{figure}

Figure \ref{fig:secure} shows the secret key bit rate $S_{AB}$ as
a function of $\alpha_0$ and $\beta_c$ for coherent pulses ($r=0$)
and for squeezed pulses ($r=0.5$). In contrast to the average gain
of information $G_{AB}$ as shown in figure \ref{fig:lossless}, the
amount of \textit{secret} information $S_{AB}$ assumes a maximum
for finite values of $\alpha_0$ and $\beta_c$. Therefore, Alice
and Bob can optimise their parameters to achieve the highest key
bit rate once they have measured the transmission loss $R^2$.
Moreover, comparing figures \ref{fig:secure}(a) and (b) one notes
that the maximum of $S_{AB}$ can be increased by using squeezed
light. This is due to the strong correlations between Bob's and
Eve's measurements in this case as discussed above.


\subsection{Lower bound on secret bit rate}

If Eve is allowed to perform any kind of measurement permitted by
quantum theory on her part of the pulse, the analysis of the
achievable secret key rate is not so straightforward. In
particular, if Alice sends a squeezed state and Bob performs a
quadrature measurement, Eve's part of the pulse is in a mixed
state in an infinite-dimensional Hilbert space due to the
entanglement created by her beam-splitter (\ref{eq:entangle}). For
such a situation the ideal measurement is not known.

I will therefore derive an upper bound on Eve's information of the
bits shared by Alice and Bob by assuming that for each bit Eve
knows (i) the basis set used, (ii) whether or not Bob has a bit
error (Eve might learn this during Alice's and Bob's error
correction procedure), and (iii) the modulus of the value measured
by Bob for his chosen quadrature component.

Let us assume that Alice sends a state in basis 1, i.e.\ state
$|\psi_0\rangle$ or $|\psi_\pi\rangle$, and that Bob measures the
real quadrature component $\beta_r$. If there is no bit error
(case I), Eve knows that her state is one of the two states
 \be
 |\chi_{I+}\rangle = {}_B\big\langle|\beta_r| \big|\psi_0\big\rangle, \quad
 |\chi_{I-}\rangle = {}_B\big\langle-|\beta_r|
 \big|\psi_\pi\big\rangle,
 \ee
where $|\pm\beta_r\rangle_B$ is the quadrature eigenstate
corresponding to $\pm\beta_r$ in Bob's Hilbert space. This occurs
with probability $(1-\delta(\beta_r))P(|\beta_r|)$, where
$\delta(\beta_r)$ is given by (\ref{eq:dbr}) and $P(|\beta_r|)$ is
the probability of finding $|\beta_r|$ in a post-selected sample
of pulses,
 \be
 P(|\beta_r|) = \frac{P_0(\beta_r)+P_0(-\beta_r)}{P(0)+P(1)}.
 \ee
In case of a bit error in Bob's detection (case II), Eve finds one
of the two states
 \be
 |\chi_{II+}\rangle = {}_B\big\langle|\beta_r| \big|\psi_\pi\big\rangle, \quad
 |\chi_{II-}\rangle = {}_B\big\langle-|\beta_r|
 \big|\psi_0\big\rangle.
 \ee
The probability for this case is $\delta(\beta_r)P(|\beta_r|)$. In
both cases, Eve is left with the task to distinguish between two
non-orthogonal states. She thus can perform an ideal orthogonal
projection measurement similar to the one discussed in section
\ref{sec:interceptideal} which minimises her bit error rates
$\delta_{I}(|\beta_r|)$ and $\delta_{II}(|\beta_r|)$,
respectively, to \cite{ban}
 \be
 \delta_{I,II}(|\beta_r|) = \frac{1}{2}\left(1-
  \sqrt{1-|\langle\chi_{I,II+}|\chi_{I,II-}\rangle|^2}
 \right).
 \ee
The collision rate $P_c$ for this kind of attack can now be
obtained by averaging over all initial states and all measurement
results \cite{norbert1}. The result is given by
 \bea
 P_c=\int_{\beta_c}^{\infty}d\beta_r\, P(|\beta_r|)\Big\{
 (1-\delta(\beta_r))\left[\delta_{I}(|\beta_r|)^2+(1-\delta_{I}(|\beta_r|))^2\right]
 & & \nonumber \\
 +\delta(\beta_r)\left[\delta_{II}(|\beta_r|)^2+(1-\delta_{II}(|\beta_r|))^2\right]
 \Big\}. & &
 \label{eq:pcideal}
 \eea

Let us now investigate the maximum secret key bit rate
(\ref{eq:sab}) using the collision rate (\ref{eq:pcideal}) as a
function of the channel loss and the amount of squeezing. To this
end $S_{AB}$ is optimised numerically with respect to $\alpha_0$
and $\beta_c$ for various values of $r$ and $R^2$.

\begin{figure}[tb]
 \centerline{
 \includegraphics[height=4.5cm]{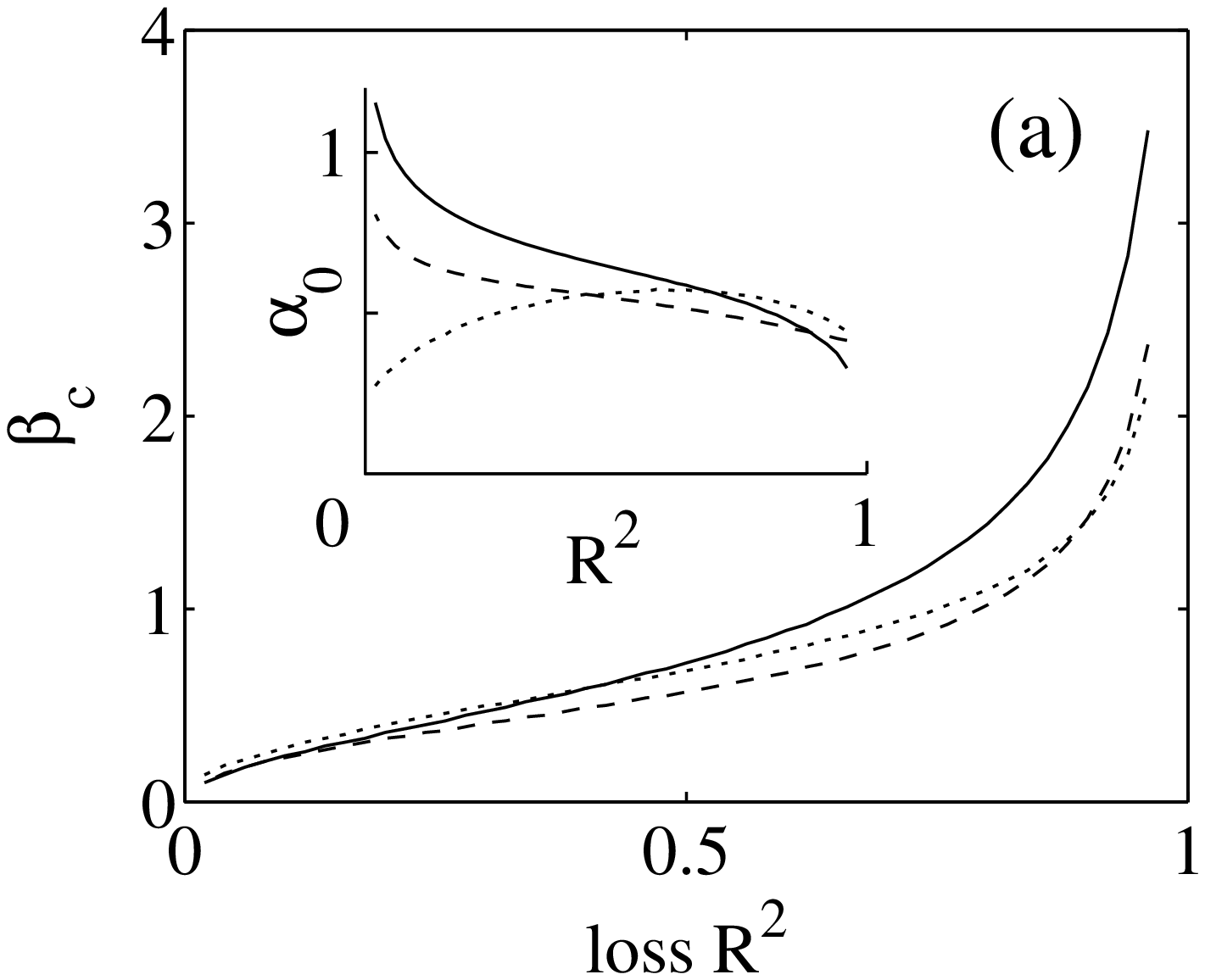}
 \includegraphics[height=4.5cm]{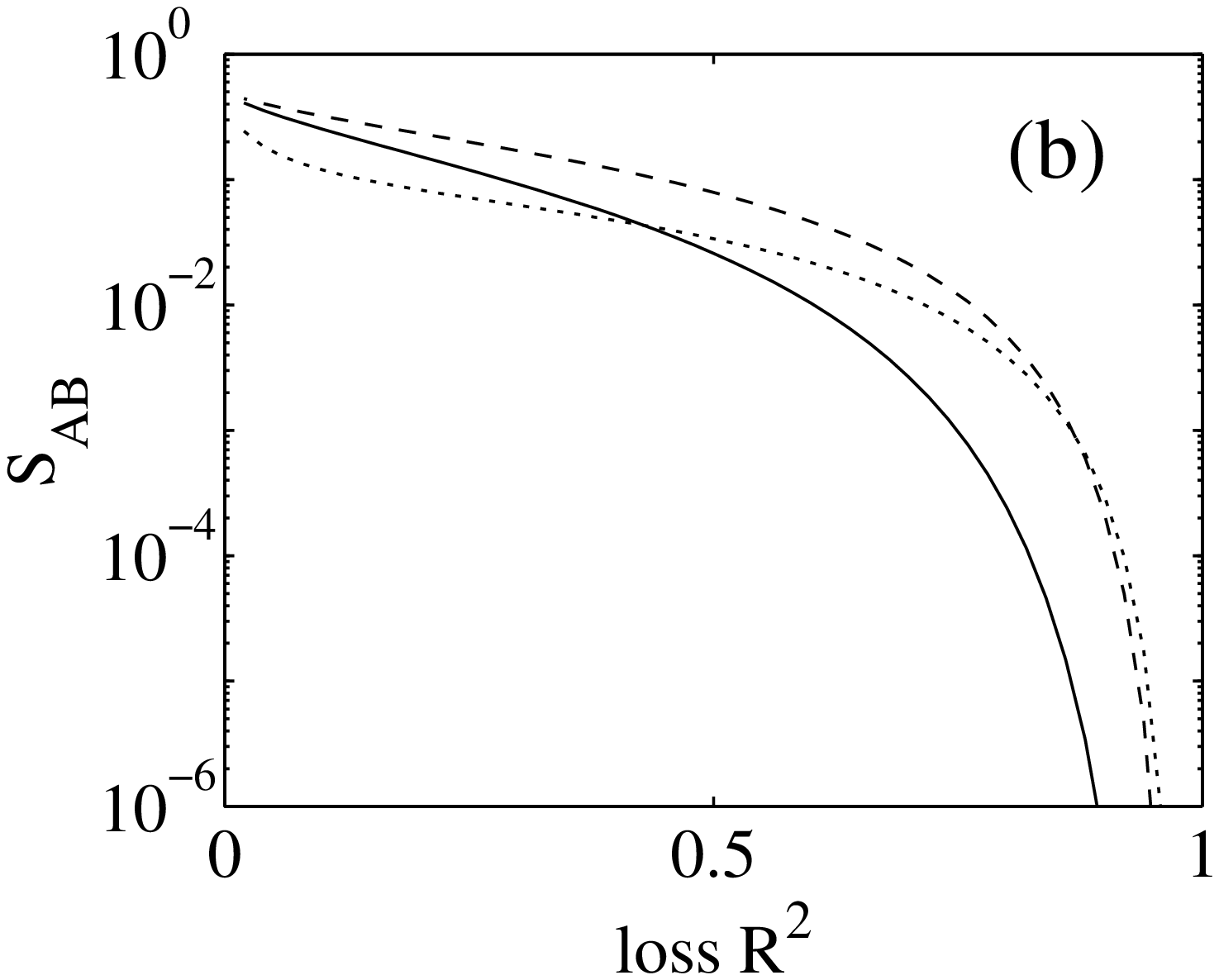}
 }
 \caption{(a) Threshold $\beta_c$ and pulse amplitude $\alpha_0$
 (inset) for maximum secret key rate $S_{AB}$ versus transmission loss
 $R^2$. (b) Corresponding values of $S_{AB}$. Solid curves are for
 $r=0$, dashed curves for $r=0.5$, dotted curves for $r=2$.
 }
 \label{fig:optloss}
\end{figure}

Figure \ref{fig:optloss} shows the optimised values of $\alpha_0$,
$\beta_c$ and $S_{AB}$ for three different values of $r$ as a
function of the loss $R^2$. Generally, $\alpha_0$ is of the order
of 1 and is slightly decreasing with increasing loss for coherent
states ($r=0$) and weakly squeezed states ($r=0.5$). For strong
squeezing ($r=2$) there is a maximum of $\alpha_0$ at around
$R^2\approx 0.6$. The ideal post-selection threshold $\beta_c$, on
the other hand, is always increasing from close to zero at low
losses to values of approximately 3 at $R^2=0.96$. The maximum of
$S_{AB}$ is close to 0.5 at low loss, decreases relatively slowly
for moderate losses and drops dramatically for high losses of
about 85-90\%. Using weakly squeezed light increases $S_{AB}$, in
particular for large values of $R^2$.

For strong squeezing and small $R^2$, the lower bound for the
secret information derived here decreases below the result for
coherent ($r=0$) states. In this limit, the entanglement between
Bob's and Eve's states becomes very strong and thus the additional
assumption (iii), i.e.\ assuming Eve knows $|\beta_r|$,
significantly enhances her detection efficiency and therefore
reduces $S_{AB}$. By contrast, in the absence of squeezing Eve's
pulse is independent of Bob's measurement result and always in one
of two possible coherent states which she can discriminate
efficiently by a two-state orthogonal projection measurement.
Therefore, the above estimate of a lower bound on the secret
information is in fact a tight bound for $r=0$, but it
over-estimates Eve's information for large squeezing.

\begin{figure}[tb]
 \centerline{
 \includegraphics[height=4.5cm]{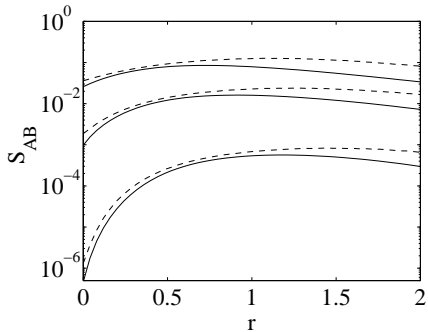}
 }
 \caption{Maximum secret key rate $S_{AB}$ versus squeezing parameter
 $r$. Solid curves show the lower bound for transmission losses of
 $R^2=0.5$, $R^2=0.75$ and $R^2=0.9$ (from top to bottom). Dashed
 lines show the corresponding results when the eavesdropper is
 restricted to quadrature measurements.
 }
 \label{fig:optsqueez}
\end{figure}

The effects of squeezing are even more apparent in figure
\ref{fig:optsqueez}, where $S_{AB}$ is plotted as a function of
$r$ for fixed loss rates. In all cases, $S_{AB}$ increases with
increasing $r$, reaches a maximum for $r\sim 1$ and decreases
again for large values of $r$ as discussed before. For 50\% loss,
$S_{AB}=0.026$ for $r=0$ and 0.085 for $r=0.72$. For 75\% loss,
$S_{AB}$ increases approximately from 0.001 to 0.016 at $r=0.92$,
and for 90\% loss from $5\times 10^{-7}$ to $6\times 10^{-4}$ at
$r=1.2$.

For comparison, the dashed curves in figure \ref{fig:optsqueez}
show $S_{AB}$ provided that Eve is restricted to quadrature
measurements as discussed in the previous section. As expected,
$S_{AB}$ is always above the lower bound shown by the
corresponding solid curves. However, it can be seen that the
secure key rate is only increased by a factor of less than two for
moderate squeezing with $r<1$. Thus the comparably simple
quadrature measurement appears to be a near-optimal eavesdropping
strategy.


\section{Summary and conclusions}
\label{sec:conclusions}

In this work, I have investigated the effects of using squeezed
pulses of light in continuous-variable quantum key distribution
with post-selection \cite{leuchs,namiki}. In this scheme, Alice
prepares the pulses in one of four squeezed states corresponding
to the bit values 0 and 1 in two non-orthogonal basis sets. Bob
uses homodyne detection to measure one of the two orthogonal
quadrature components and disregards all bits where his measured
value is below a certain post-selection threshold.

In particular, two kinds of eavesdropping strategies have been
investigated. The first is a capture-resend attack where Eve
attempts to guess correctly which of the four states was sent by
performing a measurement on the whole pulse. Linear as well as
nonlinear measurements have been discussed and the optimal
projection operator has been constructed. Using squeezed pulses
renders the scheme more robust against this kind of attack, as it
reduces Eve's chances of guessing the correct state of the pulse.
In principle, Alice and Bob can achieve a similar level of
security with coherent pulses, i.e.\ without squeezing, by
increasing the post-selection threshold but only at the cost of a
smaller key bit rate.

The second eavesdropping strategy considered in this paper is a
passive superior channel attack where Eve's interference with the
quantum communication is indistinguishable from the channel loss.
A lower bound has been derived for the secure key bit rate which
Alice and Bob can obtain by ideal error correction and privacy
amplification. Here, squeezing introduces strong correlations
between Bob's and Eve's measurement results. These correlations
allow Bob to restrict Eve's knowledge about the transmitted bits
by the post-selection process. For a given channel loss, the
secure key bit rate can be maximised by a proper choice of the
pulse amplitude and of the post-selection threshold. The
calculations show that already modest amounts of squeezing, such
as a squeezing parameter of $r=0.5$ (approximately 60\% squeezing
below shot noise), can increase the maximum secure key bit rate by
over two orders of magnitude for transmission losses of 90\%,
corresponding to transmission through 50 km of optical fibre with
losses of 0.2 dB/km.

These two classes of eavesdropping strategies are among the most
frequently discussed attacks in the literature on quantum
cryptography. The intercept-resend attack is the most `classical'
attack and one of the very few attacks which can be realised with
present-day technology, while the superior channel attack has been
shown to be the ideal eavesdropping strategy for some quantum
cryptography schemes. I have proven that the scheme discussed here
is secure against these two attacks. This, however, is no general
proof of security. In particular active eavesdropping using, for
example, an entangling cloner which has been suggested as an ideal
attack for some continuous variable crypto-schemes \cite{grangier}
is not included here and requires further analysis.

In conclusion, this work shows that squeezing is not an essential
requirement for secure quantum key distribution, in accordance
with earlier results \cite{revrecon,leuchs,namiki,grangier}.
However, it will increase the security of the scheme against
several specific eavesdropping attacks and can thereby
significantly increase the amount of secure information
transmitted per pulse.

\textbf{Acknowledgments}. This work was supported by the United
Kingdom Engineering and Physical Sciences Research Council.


\end{document}